%% file: main.tex
\documentclass[%
reprint,
amsmath,amssymb,
aps,
prl,  
superscriptaddress, 
longbibliography,
]{revtex4-1}

\usepackage{graphicx}
\usepackage{dcolumn}
\usepackage{bm}
\usepackage{graphicx} 
\usepackage[utf8]{inputenc}
\usepackage{color}
\usepackage{xspace}
\usepackage[nolist]{acronym}
\usepackage[mathlines]{lineno}
\usepackage{hyperref}
\usepackage[normalem]{ulem}
\hypersetup{colorlinks = true, allcolors = blue}

\graphicspath{{./},{Figures/}}

\begin{document}

	
\preprint{APS/123-QED}
	
\title{Determining $g_{A}/g_{V}$ with High Resolution Spectral Measurements Using an LiInSe$_2$ Bolometer}%
	
\input{author_list}

\date{\today}

\begin{abstract}

Neutrinoless Double-Beta decay (0$\nu\beta\beta$) processes sample a wide range of intermediate forbidden nuclear transitions, which may be impacted by quenching of the axial vector coupling constant (\ensuremath{g_A/g_V}\xspace), the uncertainty of which plays a pivotal role in determining the sensitivity reach of 0$\nu\beta\beta$ experiments. In this Letter, we present measurements performed on a high-resolution LiInSe$_{2}$~ bolometer in a ``source=detector'' configuration to measure the spectral shape of the 4-fold forbidden $\beta$-decay of $^{115}$In. The value of $\ensuremath{g_A/g_V}\xspace$ is determined by comparing the spectral shape of theoretical predictions to the experimental $\beta$ spectrum taking into account various simulated background components as well as a variety of detector effects.  We find evidence of quenching of $\ensuremath{g_A/g_V}\xspace$ at $>5\sigma$ with a model-dependent quenching factor of $0.655\pm0.002$ as compared to the free-nucleon value for the Interacting Shell Model.  We also measured the $^{115}$In half-life to be [$5.18\pm0.06(\text{stat.})^{+0.005}_{-0.015}(\text{sys.})]\times{10}^{14}$ yr within the Interacting Shell Model framework.  This work demonstrates the power of the bolometeric technique to perform precision nuclear physics single-$\beta$ decay measurements, which can help reduce the uncertainties in the calculation of $0\nu\beta\beta$ nuclear matrix elements.





\end{abstract}

\maketitle

\input{acronyms}


\section{Introduction}

From the first observation of single $\beta$-decay~\cite{Chadwick1914} that led W. Pauli to propose the neutrino~\cite{Pauli1930} and the subsequent efforts to develop a theory of $\beta$-decay by E. Fermi~\cite{Fermi1934} to C.S. Wu's ground-breaking work to determine the vector and axial vector form of the weak interaction~\cite{Wu1957}, the study of $\beta$-decay has been used to elucidate the hidden world of nuclear and particle physics. Modern efforts continue this legacy, using nuclear $\beta$-decay to investigate the properties of neutrino mass including its absolute scale through endpoint measurements~\cite{Aker2019,Aker2021,Aker2022}, and possible Majorana origin through searches for \OvBB~\cite{KamLANDZen2016,Alvis2019,Anton2019,Agostini2020,Adams2020,Adams2022,Armengaud2021,Azzolini2019a,Alenkov2019}.

In recent years, cryogenic bolometers have established themselves as a powerful technology in rare event searches for \OvBB \cite{Adams2020,Adams2022,Armengaud2021,Azzolini2019a,Agostini2020,Alvis2019,Alenkov2019}, direct Dark Matter detection \cite{Abdelhameed2019,Alkhatib2021,Agnese2019}, and more \cite{Beaulieu2021,Alkhatib2021a,Pyle2015,Poda2021,DeMarcillac2003}. Such detectors operate at milli-kelvin temperatures and measure energy deposition events by converting phonons into a temperature increase within a sensitive thermistor. Bolometers benefit from excellent energy resolution, high electron containment efficiencies, low energy trigger thresholds, and strong particle-ID capabilities when equipped with a dual heat/light or heat/ionization readout~\cite{Poda2021,Huang2021,Azzolini2019a}.  Additionally, the ability to operate nearly any crystalline material as a bolometer provides practical means to study a very wide range of long-lived nuclear processes for which sufficient quantities of isotope may be procured and grown into crystalline form. 

As pointed out by \cite{Barea2013}, theoretical calculations of the nuclear physics contributions to the \OvBB half-life have often assumed an axial-to-vector coupling ratio equal to that of the free neutron,~\gAgV = 1.276~\cite{Mund2013,Markisch2019}, though it is common to use a quenched value to obtain agreement with observed single-$\beta$ transition rates ~\cite{Kumar2016,Chou1993,Brown1988,Martinez-Pinedo1996}. The exact impact on \OvBB will depend on the underlying physics of axial quenching \cite{Suhonen2017a}; recently Ref.~\cite{Gysbers2019} provided evidence that the inclusion of two-nucleon currents and additional correlations may provide an explanation within light ($A\leq14$) nuclei and certain super-allowed heavy nuclei $\beta$-decay transitions. Axial quenching creates a significant uncertainty in the interpretation of any \OvBB search when converting isotope-specific half-lives back to the underlying physics of interest \cite{Dolinski2019}, on top of the existing spread in the value of calculated Nuclear Matrix Elements (NMEs) for \OvBB isotopes \cite{Engel2017}. 

As was proposed in~\cite{Haaranen2016}, the \emph{shape} of highly-forbidden $\beta$-decay spectra can be very sensitive to \gAgV, and studying such decays of nuclei with mass around $A\sim100$ could shed light on axial quenching in a similar nuclear environment as those found in \OvBB decays. This analysis technique could also have applications in explaining reactor flux anomalies through examination of 1st-order forbidden $\beta$-decay transitions~\cite{Hayen2019}. This spectral shape technique was used for the first time in~\cite{Alessandrello1994}, where the experimental data from a CdWO$_4$ scintillation detector~\cite{Belli2007} were compared to theoretical spectra in order to extract a value for $g_A$ in the range of 0.90--0.93. More recently, COBRA has applied this spectral shape approach to their data of CdZnTe detectors in order to obtain a range for $g_A$ between 0.92 and 0.96 depending on the theoretical models used~\cite{Bodenstein-Dresler2020}.
In this Letter, we make a precision $\beta$-decay spectral shape measurement using a high-resolution ``source=detector'' bolometer. In particular, we study the 4-fold forbidden $\beta$-decay of \mbox{$\isoIn\rightarrow{}^{115}{\rm Sn}$} with \mbox{$Q_\beta=497.489$\,keV}~\cite{Wang2021} with the most recent previously measured half-life of ($4.41 \pm 0.25) \times 10^{14}$ years~\cite{Pfeiffer1979}. This decay occurs in a mass range relevant to \OvBB isotopes of interest and provides a benchmark to test whether many-body nuclear calculations are capable of simultaneously explaining the $\beta$-decay spectral shape and rate.  Recently, interest has been growing to measure this particular \isoIn decay mode by examining an In$_2$O$_3$ bolometer in order to provide a measurement of \gAgV~\cite{Celi2022}. 
Here we use a \LIS crystal with a natural abundance of \isoIn (95.72 \% \cite{Meija2016}), to evaluate \gAgV for leading nuclear models, and make the most precise measurement of the \isoIn half-life to date.

\section{Methods}

\begin{figure}
\centering
\includegraphics[width=.45\textwidth]{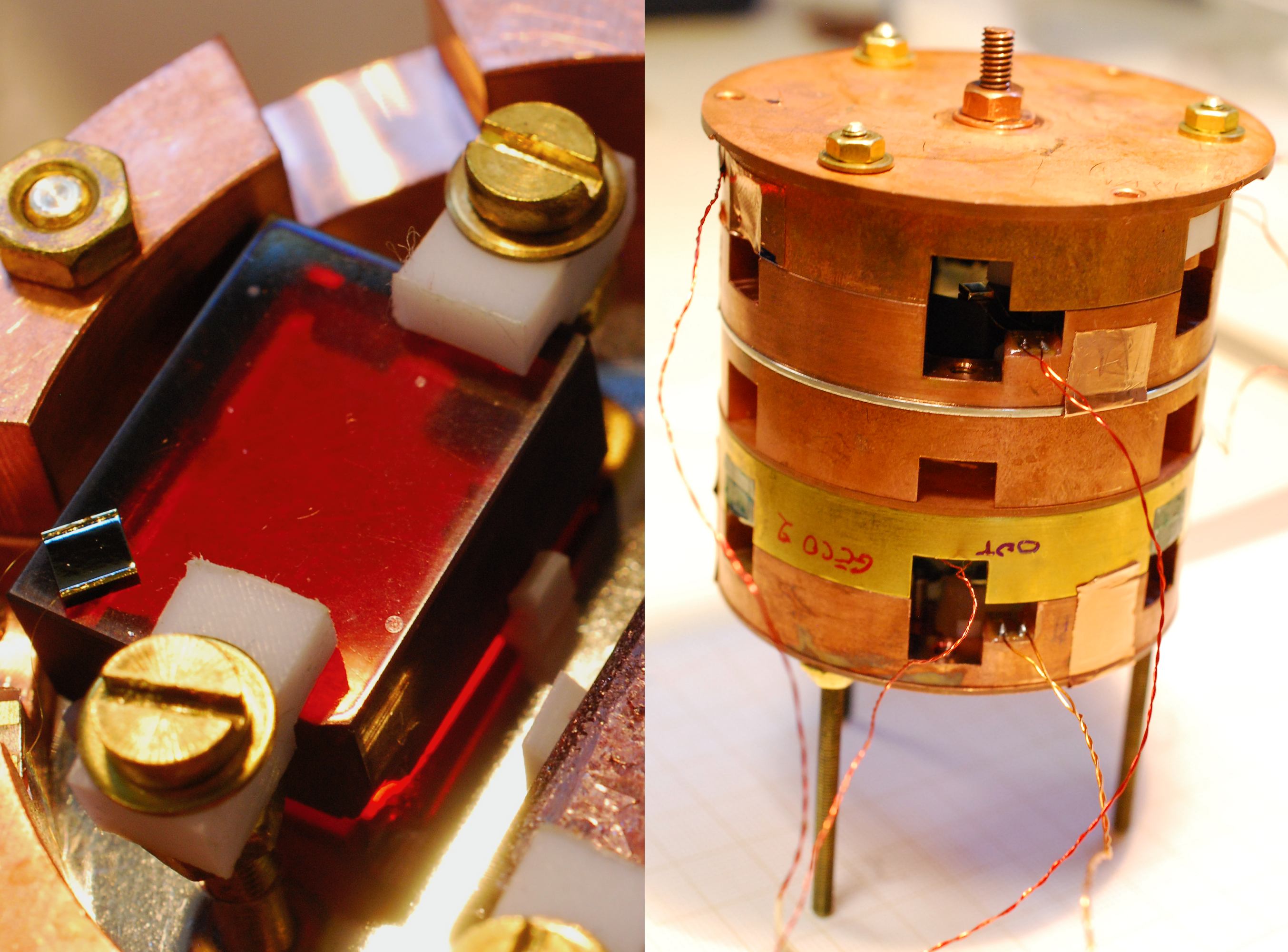}
\caption{(Left) Photo of the \LIS bolometer with an NTD thermistor attached to the crystal. (Right) The combined detector setup in a tower configuration with two pairs of bolometers stacked in two stages. The light detector is placed above each ''stage'' of the tower for maximum photon absorption}
\label{fig:BolometersPhoto}
\end{figure}

The \LIS crystal was grown by RMD Inc.~\footnote{44 Hunt Street, Watertown, MA 02472, email: info@rmdinc.com, website: \href{https://www.rmdinc.com/}{rmdinc.com}} using the vertical Bridgman process~\cite{Isaenko2005,Tower2020}.
The crystal was enriched in $^6$Li to 95\% for potential use as a neutron detector \cite{Bell2015,Tupitsyn2012}, however, that analysis is beyond the scope of this work and does not affect the $\beta$-decay analysis. The \LIS crystal was instrumented with a \NTD thermistor \cite{Haller1984}, and installed inside a cryostat at IJCLab (ex. CSNSM) in Orsay, France \cite{Danevich2018}, see Fig.~\ref{fig:BolometersPhoto}. The \LIS scintillation signal was monitored by a separate Neganov-Trofimov-Luke Ge \LD\cite{Novati2019}, which allowed us to perform particle identification and pile-up rejection. 42.2 g$\cdot$days of data was collected over two weeks, with the pertinent measured/derived experimental parameters summarized in Table ~\ref{table:ExperimentParams}. 

\begin{table}[]
	\caption{Experimental parameters of the \LIS crystal during the October-November 2017 data runs.}
	\label{table:ExperimentParams}
	\begin{tabular}{|c|c|}
		\hline
		Detector Parameter & LiInSe$_{2}$ Crystal \\ \hline \hline
		\multicolumn{1}{|c|}{Crystal Dimensions}       & $1.3\times1.6\times0.7$ cm      \\ \hline			
		\multicolumn{1}{|c|}{Total Crystal Mass}       & 10.3 grams       \\ \hline	
		\multicolumn{1}{|c|}{Effective \isoIn Mass}       & 4.1 grams       \\ \hline		
		\multicolumn{1}{|c|}{Noise Level}        & 1.1 keV (1$\sigma$)   \\ \hline
		\multicolumn{1}{|c|}{Avg. Energy Resolution}       & 2.4 keV (1$\sigma$)       \\ \hline
		\multicolumn{1}{|c|}{100 \% Trigger Threshold} & 20.0 keV           \\ \hline
		\multicolumn{1}{|c|}{Analysis Threshold} & 160 keV             \\ \hline
		\multicolumn{1}{|c|}{Containment Eff.}        & 96.6\% @ 497 keV \\ \hline
		\multicolumn{1}{|c|}{Data Selection Cut Eff.}        & 47.6(2)\% ($160-500$ keV) \\ \hline
		\multicolumn{1}{|c|}{Livetime Fraction}        & 52.54(8)\% \\ \hline
		\multicolumn{1}{|c|}{Total Exposure}     & 39.7 g$\cdot$days \\ \hline	
	\end{tabular}
\end{table}

The data was processed using the \apollo/\diana software developed by the CUORE~\cite{Alduino2016}/CUPID-0~\cite{Azzolini2018}/CUORICINO~\cite{Andreotti2011} collaborations. Events are triggered with the \OT ~\cite{DiDomizio2011} and processed following a procedure similar to \cite{Alduino2016,Adams2020}.
The trigger threshold was determined by injecting a series of low energy pulses through the attached Joule heater~\cite{Andreotti2012}, achieving $\sim100$\% trigger efficiency above 20\,keV. The \LIS detector is calibrated with a set of dedicated runs with a $^{133}$Ba source using the four most prominent $\gamma$ peaks in the energy range 250--400\,keV.

\begin{figure}
\centering
\includegraphics[width=.48\textwidth]{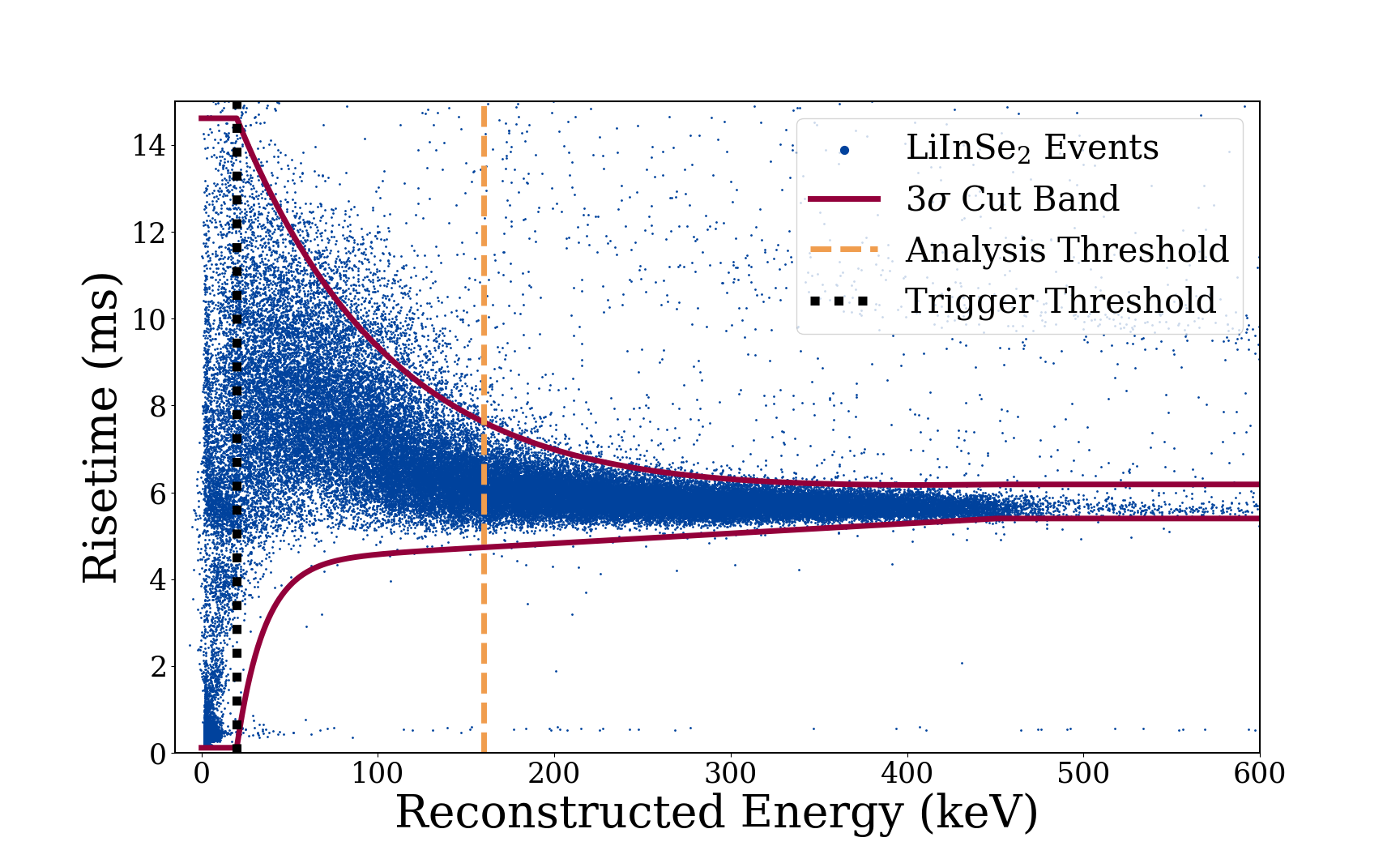}
\caption{\LIS detector events with $3\sigma$ cut bands, analysis and trigger thresholds superimposed. The corresponding rise times were collected in 10 keV energy bins running between 20-450 keV and then each bin of rise times were individually fit to a Gaussian. The cut band was then defined by interpolating between the individual 3$\sigma$ profiles cuts as a function of energy. Outside of the 20--450 keV energy range, the cut values were kept constant due to large uncertainties in the profile fit parameters as a result of non-Gaussian parameter distributions or low statistics at the low/high energy ranges respectively.}
\label{fig:LISDQcuts}
\end{figure}

The internal \isoIn decay on its own results in an expected event rate of $\approx1.2$\ Hz in the 10.3 gram \LIS detector, which means that internal event pile-up is expected to be a significant background. The recovery time after an event is $\sim$200\,ms, and the event window around each event includes 100\,ms before the trigger and 500\,ms after. Together, these lead to a significant paralyzable deadtime.  The faster response time of the \LD allows us to efficiently tag and remove these pile-up events that might otherwise slip through the \LIS data quality cuts (see Fig.~\ref{fig:LISDQcuts}), along with tagging $\alpha$ events through particle-identification via event-by-event light-yield cuts. 

\begin{figure}
\includegraphics[width=.48\textwidth]{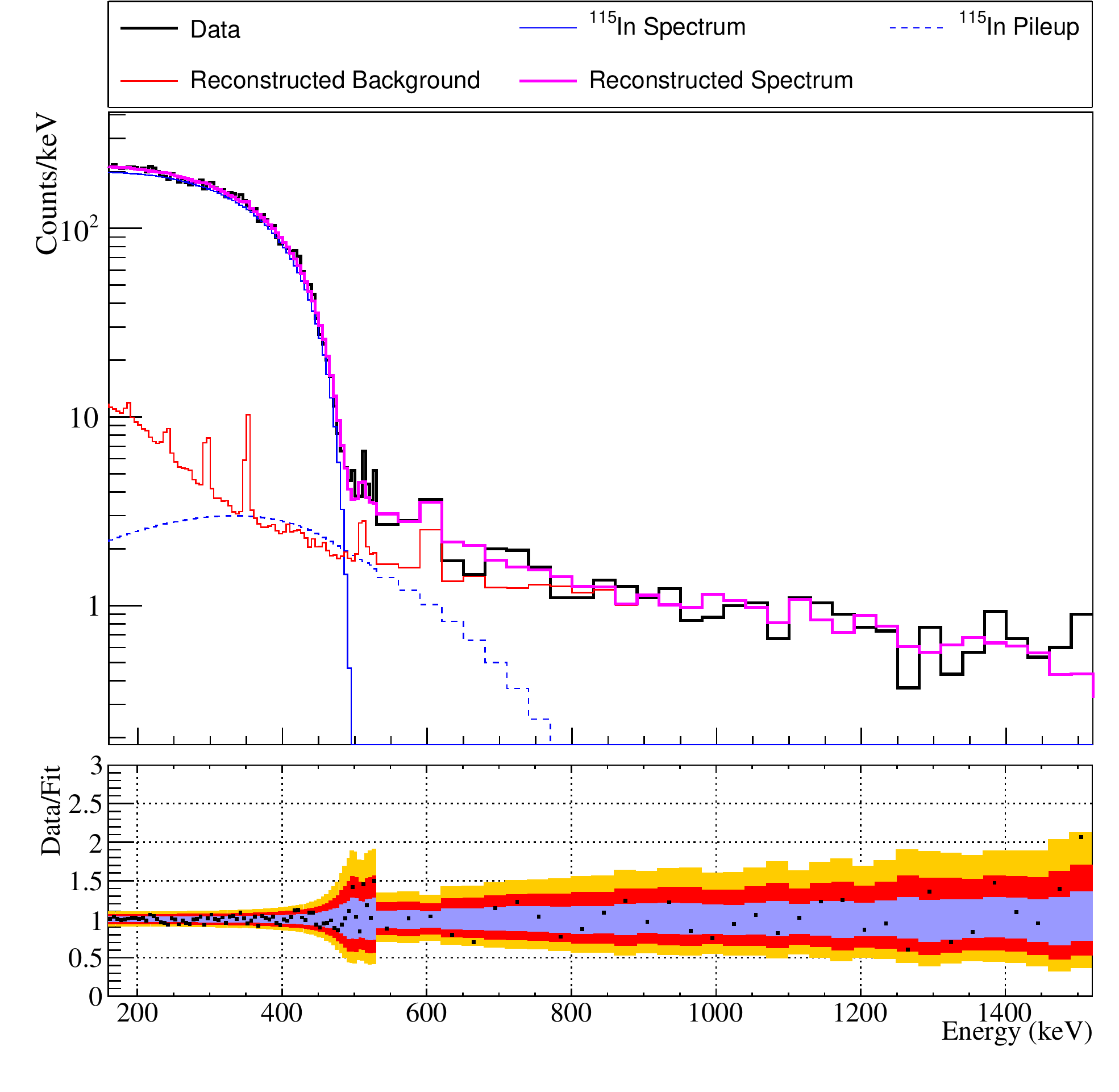}
\caption{Spectral fit to the collected \LIS spectrum over the region 160--1520 keV.  Component normalizations and the \isoIn spectral shape correspond to the best-fit values for the Interacting Shell Model (\ISM) exhibiting a $\chi^2$ value of 160 with 101 degrees of freedom. Fits to the Microscopic Quasi-particle-Phonon Model (\MQPM) and Interacting Boson Model (\IBM) result in similar reconstructions. The bottom panel displays Data/Fit ratios for the reconstruction, along with 1$\sigma$ (purple), 2$\sigma$ (red) and 3$\sigma$ (yellow) fit credibility regions. The spectrum is binned by 5 keV up until 530 keV and by 30 keV above 530 keV in order to maintain reasonable statistics per bin above the \isoIn endpoint.}
\label{fig:FitSpectrum}
\end{figure}

In order to filter out spurious events from $^{115}$In $\beta^-$ events, a series of loose pulse shape cuts were employed to filter out electrical glitch and badly reconstructed events. Then a rise time pulse quality cut (see Fig.~ \ref{fig:LISDQcuts}) was defined by a 3$\sigma$ cut band determined by fitting the resulting pulse shape variable profiles across each energy bin. We also employ a coincidence cut that enforces a single-event criterion. We require that an event is included in the final spectrum if it appears on both the \LIS and the \LD detectors within 20 ms and no other events are recorded on the \LIS detector within a broader 600 ms window. Over the region of 160--500 keV, we find a cut efficiency of ($47.6\pm0.2$)\%, dominated by the \LD single-event criterion. The 160 keV threshold was selected as the lowest energy where multiple event pile up was well handled by the autoconvolution background component. The resulting events that pass all the above cuts are then compiled into the input \LIS spectrum as shown in Fig.~\ref{fig:FitSpectrum}. 

To extract \gAgV from the measured \LIS spectrum, we follow a procedure similar to~\cite{Adams2021,Azzolini2019,Armengaud2020,Alduino2017} and decompose  it into various components: a model-dependent signal component from the $\beta$-decay of \isoIn which will depend on \gAgV, an untagged pile-up component, and other radioactive background contributions. The fit is implemented using the Bayesian Analysis Toolkit package \cite{BAT2009}, which implements a Markov Chain Monte-Carlo (MCMC) to sample the full joint posterior.  We perform this decomposition on the spectrum in Fig.~\ref{fig:FitSpectrum}, which has a binning of 5/30\,keV below/above 530\,keV respectively up until the analysis cut-off at 1520 keV. This binning scheme allows for the fitting of as many broad spectral/peak features as possible present in the experimental data while maintaining the highest possible statistics per bin in the region beyond 530 keV.  Despite the low trigger threshold of the \LIS crystal, we implement an analysis threshold of 160 keV to avoid low-energy pile-up events which are difficult to separate in time and can distort the spectrum.

To implement the MCMC, we define our binned likelihood as: 
\begin{equation}
    \mathcal{L}=\prod_i \text{Pois}\left(k_i;\sum_{j} a_j \lambda_{ij}  \right), 
\end{equation}
enumerating bins by $i$ and fitted components by $j$.  Here, $k_i$ is the number of observed counts within a given bin, $\lambda_{ij}$ is the normalized density of the $j^{\text{th}}$ component within the $i^{\text{th}}$ bin, and $a_j$ are the fitted normalizations for the different components.  The densities $\lambda$ corresponding to \isoIn are \gAgV-dependent.

A numerical calculations for the structure of \isoIn are performed using \ISM \cite{Iwata2016,Menendez2009,Horoi2016}, \IBM \cite{Barea2015} and \MQPM \cite{Toivanen1998}. The resulting $\beta$-decay spectrum is generated as a function of energy for each of these structural models taking \gAgV as an input.
We generate a library of 200 discrete $\beta$-decay spectra for \gAgV uniformly spaced across the range $0.6<\gAgV<1.3$ and then perform an interpolation for the spectral shape for \gAgV values not in our library.  Each \isoIn spectrum is then convolved with an energy-dependent detector response function to account for energy losses as well as shifts in the spectral shape from $\beta$-particles that escape the absorber. This is calculated through a Geant4 simulation \cite{Allison2016} that only simulates the \LIS crystal and the copper plate it rests on.  These simulations find that 96.6\% of electrons at the $\beta$-decay endpoint will be fully contained within the detector, which represents the minimum containment efficiency over the \isoIn spectrum. Background component spectra are obtained by simulating in Geant4 various possible radiogenic contaminations including from daughter nuclei on various detector/cryostat components, for example the copper cryostat cold plates and lead shielding surrounding the detector. In total, we simulated the $\gamma$/$\beta$ spectra stemming from $^{238}$U/$^{232}$Th decay chains as well as $^{60}$Co, and $^{40}$K decays present uniformly throughout the \LIS detector, to simulate possible contamination of the various cyrostat components particularly the copper plate near the detector, and from external environmental sources. In addition, we simulated a separate background contribution coming solely from possible surface contaminations of the \LIS crystal. All these input spectra components were then convolved with the detectors' measured energy resolution and binned into the same binning scheme as the data before their use as a potential component of the MCMC fit. The inclusion of a pile-up component (the autoconvolution of the \isoIn $\beta$-spectrum) was designed to account for the inability to separate events which occur too closely in time and could then be mis-reconstructed as a single higher energy event.

The final MCMC fit only included the four most-dominant background components: 1/2) internal crystal contamination stemming from the $^{238}$U decay chains and $^{60}$Co decays, 3) $^{232}$Th decay chain events on the copper plate underneath the \LIS crystal, and 4) $^{232}$Th decay chain events from external sources mostly in the form of $\gamma$s. Contamination from $\alpha$ surface backgrounds can be ignored, thanks to the strong pulse shape and coincidence cuts that were applied to the collected data, resulting in predominantly bulk $\gamma$ backgrounds. All other simulated background components were found to have only a negligible effect on the final fit parameters. This results in a satisfactory description of background features in the collected spectrum without introducing degeneracies in the fit from additional components which may not be differentiated with available data. We perform a separate fit for each nuclear model tested, and apply uniform priors to the normalizations of each fitted component within the regions of \gAgV discussed below.

\section{Discussion}

For all three nuclear models examined, the likelihood function within the fit is bi-modal with respect to \gAgV, exhibiting a local minimum both at low-\gAgV values below 0.95, and at high-\gAgV values above 1.05. Fits arising from the high-\gAgV minimum result in a poor match to the observed spectral shape, with decreases in log-likelihood as compared to the low-\gAgV minimum of at least 65 (IBM), 90 (MQPM) and 118 (ISM).  Despite resulting in an overall worse fit, the high-\gAgV fit minima are still sufficiently favored that without a restricted prior, the MCMC chain will take an unreasonably long time to achieve convergence. In order to ensure a good convergence of the MCMC chain about the global minimum while avoiding numerical instabilities, we restrict ourselves to a uniform prior on \gAgV $\in [0.6, 1.0]$.

We extract the best-fit values from the maximum a posteriori point (which we will refer to as the ``best-fit'' values), along with Bayesian Credibility Regions (BCRs) for parameters of interest pertaining to the \isoIn decay rate and value of \gAgV. We marginalize over all background component normalizations as nuisance parameters; all three fits result in compatible contributions from each of the included background components.  The best-fit values for \gAgV along with the central $1\sigma$ BCRs arising from the fits are summarized in Table~\ref{tab:BestFitValues}.  Unsurprisingly, the various nuclear calculations prefer different values of \gAgV, however all models strongly reject the free-nucleon value of \gAgV$=1.276$ at $>5\sigma$ as determined by the $\Delta \log \mathcal{L}$ between the best-fit values and the free-nucleon value, assuming Wilk's theorem \cite{Wilks1938}.  

\begin{table}[hb]
\centering
\caption{Fit results for each of the three nuclear models considered.  For the parameters of interest of \gAgV and $T_{1/2}$ for \isoIn, we quote the best fit value with uncertainty given by the width of the central 68\% Bayesian credibility interval, along with the reduced-$\chi^2$ value for the best-fit reconstruction.}
\label{tab:BestFitValues}

\begin{tabular}{|c|c|c|c|}
\hline
Model   & \gAgV   & $T_{1/2}$\,($10^{14}$\,yr) & Reduced $\chi^2$ \\ \hline 
\hline
ISM & $0.830 \pm 0.002$ & $5.177\pm0.060$ & 1.58 \\
IBM & $0.845 \pm 0.006$ & $5.031\pm0.065$ & 1.50 \\
MQPM & $0.936 \pm 0.003$ &  $5.222\pm0.061$ & 1.60 \\
\hline
Pfeiffer et al. \cite{Pfeiffer1979} & & $4.41\pm0.25$ &\\
\hline
Watt and Glover \cite{Watt1962} & & $5.1\pm0.4$ &\\
\hline
Beard and Kelly \cite{Beard1961} & & $6.9\pm1.5$ &\\
\hline
\end{tabular}
\end{table}

\begin{figure}[h!]
\centering
\includegraphics[width=.48\textwidth]{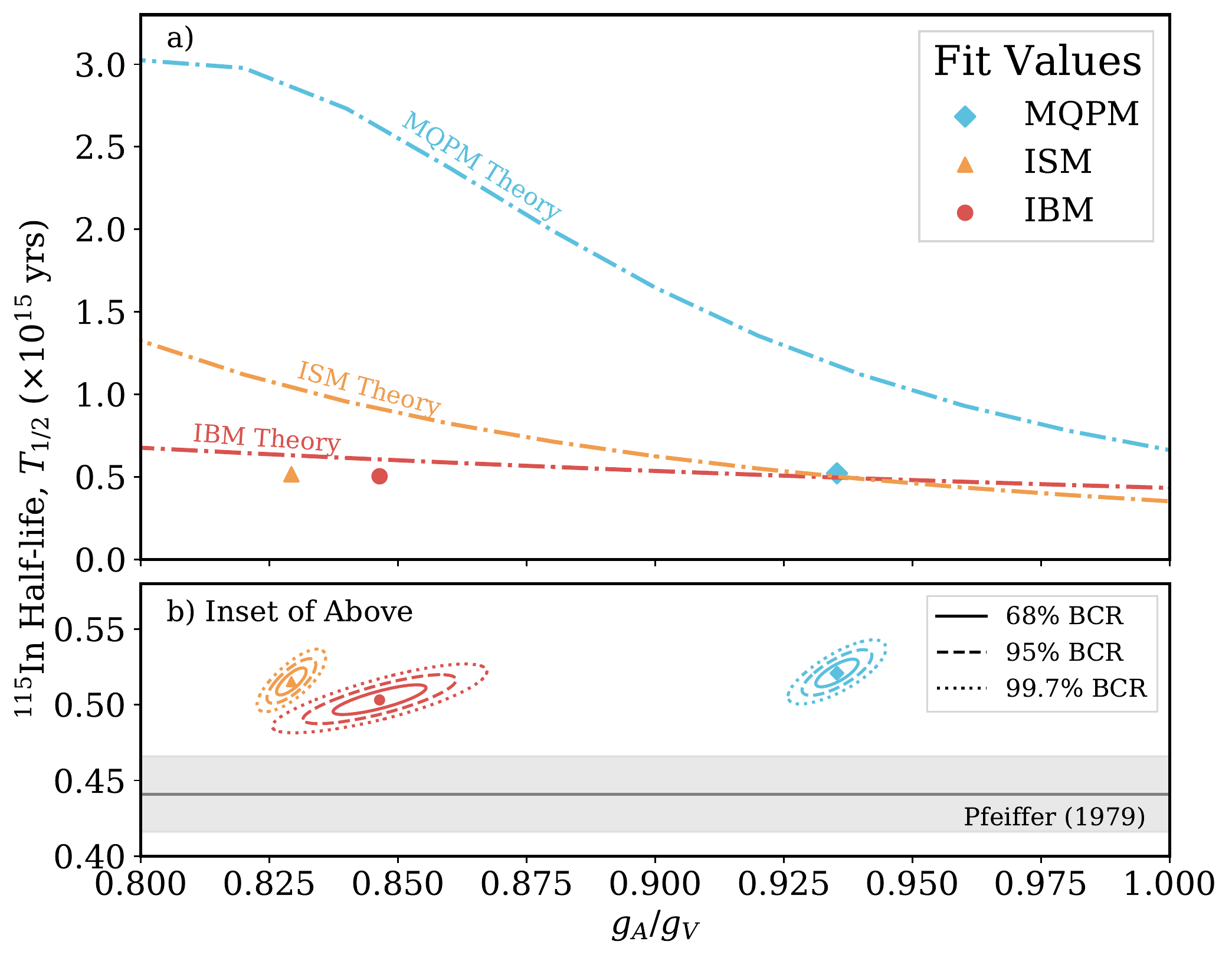}
\caption{Top: Half-lives verses \gAgV theory curves (dash-dot lines), assuming a conserved vector current \cite{Feynman1958}, for \isoIn as well as the best fit half-lives and \gAgV values (markers) resulting from the spectral-shape fits for the IBM (red), ISM (yellow), and MQPM (cyan) models considered in this Letter.
Bottom: Inset of above, focused around the experimentally determined half-life values for \isoIn. Contours about the best fit values represent the joint two-dimensional Bayesian credibility regions produced from the fit posteriors and only include statistical uncertainties. The previous half-life measurement from \cite{Pfeiffer1979} is shown in gray with $1\sigma$ uncertainty (other measurements omitted for clarity).}
\label{fig:FitValues}
\end{figure}

Additionally, using the normalization of the \isoIn component, we can extract the value of the half-life $T_{1/2}(^{115}\mathrm{In}) = [5.18\pm0.06(\text{stat.})^{+0.005}_{-0.015}(\text{sys.})]\times{10}^{14}$ years. Here we quote the best-fit value arising from the ISM model fit, with statistical uncertainty determined by the width of the $1\sigma$ central BCR with negligible contributions from uncertainties in the cut and live-time efficiencies which are propagated on top of the fitted \isoIn normalization. We choose to quote the spread in half-life with respect to the IBM and MQPM best-fit values (shown in Table~\ref{tab:BestFitValues}) as a systematic uncertainty. This is slower by 3$\sigma$ with respect to the measurement within \cite{Pfeiffer1979}, but falls within 2$\sigma$ of the older, less precise measurements \cite{Watt1962,Beard1961}.  Figure~\ref{fig:FitValues}b) displays the joint 2-dimensional Bayesian credibility regions for \gAgV and $T_{1/2}$ for each fitted nuclear model, along with the best-fit points.

Each of the nuclear models calculations discussed in this letter are able to simultaneously calculate the $T_{1/2}$ as a function of \gAgV values~\cite{Haaranen2017} as shown by the dash-doted lines in Figure~\ref{fig:FitValues}. In our analysis, our best fit values for the half-life given by each model overestimates the half-lives by factors of 1.2 (IBM), 2.2 (MQPM), and 2.0 (ISM) compared to~\cite{Pfeiffer1979}, and simultaneously does not fall upon one of the theory curves. This suggests that quenching-dependent calculations that we used are not yet able to simultaneously match the spectral shape and decay rate in \isoIn. It is worth noting that the half-life in~\cite{Pfeiffer1979} is similarly incapable of simultaneously matching the spectral shape and decay rate. 

Previous work with COBRA $^{113}$Cd data has shown that the tension between the independently measured half-life and the quenched \gAgV values extracted from the spectral shape analysis can be relaxed via the introduction of a small relativistic nuclear matrix element correction that affects the spectral shape due to the enforcement of the conserved vector current assumption~\cite{Kostensalo2021}. Additionally due to the closeness of our results with the measurements presented in~\cite{Watt1962,Beard1961}, we do not present any conclusion regarding the accuracy of any single nuclear model presented here. This letter seeks to showcase the ability of this technique to simultaneously provide two additional experimental cross checks to any nuclear calculation model, namely spectral shape and half-life, on any provided nuclear model able to address highly forbidden nuclear $\beta$-decays.

\section{Conclusion}

From these data, it is clear that the value of \gAgV that governs this highly forbidden decay process is quenched by approximately 0.65--0.75 compared to the decay of the free neutron. Interestingly, for each of the three nuclear models examined there is strong disagreement between the measured half-life from \cite{Pfeiffer1979} and the predicted half-life value for the favored value of \gAgV calculated from spectral shape analysis. This tension could point to possible issues with regards to the many-body approaches and Hamiltonians used in the various calculation frameworks. At the same time, our better agreement with the older measurements of ~\cite{Watt1962,Beard1961} may point to additional systematics effects that could play a vital role in the determination of any half-life measurement/calculation. 

This measurement shows the utility of cryogenic bolometers for precision studies across multiple energy bins to test various spectral shapes that stem from rare/forbidden nuclear processes.  Further developments in cryogenic detectors which exhibit faster timing resolution, such as those using TESs for heat and/or light readout, would provide better separation of low-energy pile-up events and could offer even better energy resolutions than the NTDs used in this experiment~\cite{Armatol2021, Huang2021}. Coupled with further improvements in the theory calculations of the nuclear matrix elements~\cite{Kumar2020,Kumar2021}, this would allow for future studies of \isoIn and other candidate isotopes for such as $^{113}$Cd~\cite{Kostensalo2021} (for an expanded list see~\cite{Ejiri2019}) further increasing the sensitivity to \gAgV and opening the door to reducing this source of uncertainty on the nuclear matrix elements utilized by \OvBB experiments in their current and projected sensitivity limits.

\input{Acknowledgments}
\bibliography{main}

\end{document}

%% file: author_list.tex
\author{A.F.~Leder}
\email{aleder@berkeley.edu}
\affiliation{Massachusetts Institute of Technology, 77 Massachusetts Ave. Cambridge, MA 02139, USA}
\affiliation{Department of Nuclear Engineering, University of California - Berkeley, 2521 Hearst Ave, Berkeley, CA 94709, USA}

\author{D.~Mayer}
\email{dmayer@mit.edu}
\affiliation{Massachusetts Institute of Technology, 77 Massachusetts Ave. Cambridge, MA 02139, USA}
 
\author{J.~L.~Ouellet}
\affiliation{Massachusetts Institute of Technology, 77 Massachusetts Ave. Cambridge, MA 02139, USA}

\author{F.~A.~Danevich}
\affiliation{Institute for Nuclear Research of NASU, Kyiv 03028, Ukraine}

\author{L.~Dumoulin}
\affiliation{Universit\'{e} Paris-Saclay, CNRS/IN2P3, IJCLab, 91405 Orsay, France}

\author{A.~Giuliani}
\affiliation{Universit\'{e} Paris-Saclay, CNRS/IN2P3, IJCLab, 91405 Orsay, France}

\author{J.~Kostensalo} 
\affiliation{Natural Resources Institute Finland, Yliopistokatu 6B, FI-80100 Joensuu, Finland}
\author{J.~Kotila}
\affiliation{Department of Physics, University of Jyv\"askyl\"a, P.O. Box 35, FI-40014 Jyv\"askyl\"a, Finland}
\affiliation{Finnish Institute for Educational Research, University of Jyv\"{a}skyl\"{a}, P.O. Box 35, FI-40014 Jyv\"{a}skyl\"{a}, Finland}
\affiliation{Center for Theoretical Physics, Sloane Physics Laboratory Yale University, New Haven, Connecticut 06520-8120, USA}
\author{P.~de Marcillac}
\affiliation{Universit\'{e} Paris-Saclay, CNRS/IN2P3, IJCLab, 91405 Orsay, France}
\author{C.~Nones}
\affiliation{Commissariat \'{a} l'\'{E}nergie Atomique (CEA)- Saclay, 91191 Gif-sur-Yvette, France}
\author{V.~Novati}
\affiliation{Universit\'{e} Paris-Saclay, CNRS/IN2P3, IJCLab, 91405 Orsay, France}
\author{E.~Olivieri}
\affiliation{Universit\'{e} Paris-Saclay, CNRS/IN2P3, IJCLab, 91405 Orsay, France}
\author{D.~Poda}
\affiliation{Universit\'{e} Paris-Saclay, CNRS/IN2P3, IJCLab, 91405 Orsay, France}
\author{J.~Suhonen}
\affiliation{Department of Physics, University of Jyv\"askyl\"a, P.O. Box 35, FI-40014 Jyv\"askyl\"a, Finland}
\author{V.I.~Tretyak}
\affiliation{Institute for Nuclear Research of NASU, Kyiv 03028, Ukraine}
\author{L.~Winslow} 
\email{lwinslow@mit.edu}
\affiliation{Massachusetts Institute of Technology, 77 Massachusetts Ave. Cambridge, MA 02139, USA}
\author{A.~ Zolotarova}
\affiliation{Commissariat \'{a} l'\'{E}nergie Atomique (CEA)- Saclay, 91191 Gif-sur-Yvette, France}

%% file: acronyms.tex
\begin{acronym}
\acro{CUORE}{Cryogenic Underground Observatory for Rare Events}
\acro{NDBD}[0$\nu\beta\beta$]{Neutrinoless Double-Beta decay}
\acro{SM}{Standard Model}
\acro{NTD}{Neutron Transmutation Doped}
\acro{FWHM}{full-width half-max}
\acro{MC}{Monte Carlo}
\acro{MCMC}{Markov-Chain Monte Carlo}
\acro{ROI}{region of interest}
\acro{NME}{Nuclear Matrix Element}
\acro{OT}{Optimum Trigger}
\acro{LD}{light detector}
\acro{IBM}{Interacting Boson Model}
\acro{ISM}{Interacting Shell Model}
\acro{MQPM}{Microscopic Quasiparticle-Phonon Model}
\acro{ML}{maximum likelihood}
\acro{BCR}{Bayesian Credibility Region}
\acro{MAP}{maximum a posteriori}
\end{acronym}

\newcommand{\OvBB}{\ac{NDBD}\xspace}
\newcommand{\TwovBB}{2$\nu\beta\beta$\xspace}
\newcommand{\gA}{\ensuremath{g_A^{\rm eff}}\xspace}
\newcommand{\gAfree}{\ensuremath{g_A^{\rm free}}\xspace}
\newcommand{\gAgV}{\ensuremath{g_A/g_V}\xspace}
\newcommand{\altgAgV}{\ensuremath{\frac{g_A}{g_V}}\xspace}
\newcommand{\LIS}{LiInSe$_{2}$\xspace}
\newcommand{\GaSe}{GaSe\xspace}
\newcommand{\In}{\isoIn}
\newcommand{\isoIn}{\ensuremath{{}^{115}{\rm In}}\xspace}
\newcommand{\LMO}[0]{Li$_{2}$MoO$_4$\xspace}
\newcommand{\THalfNu}[0]{$T^{0\nu}_\frac{1}{2}$\xspace}
\newcommand{\THalfLIS}[0]{$T^{\text{In-115}}_\frac{1}{2}$\xspace}
\newcommand{\mbb}{\ensuremath{m_{\beta\beta}}}
\newcommand{\StanLP}[0]{$\mathcal{L}$\xspace}
\newcommand{\diana}{D\textsc{iana}\xspace}
\newcommand{\apollo}{A\textsc{pollo}\xspace}
\newcommand{\OT}{\ac{OT}\xspace}
\newcommand{\LD}{\ac{LD}\xspace}
\newcommand{\NTD}{\ac{NTD}\xspace}
\newcommand{\Alex}[1]{{\color{red}{Alex: #1}}}
\newcommand{\Jon}[1]{{\color{blue}{Jon: #1}}}
\newcommand{\LW}[1]{{\color{green}{Lindley: #1}}}
\newcommand{\Daniel}[1]{{\color{magenta}{Daniel: #1}}}
\newcommand{\NME}{\ac{NME}\xspace}
\newcommand{\ISM}{\ac{ISM}\xspace}
\newcommand{\IBM}{\ac{IBM}\xspace}
\newcommand{\MQPM}{\ac{MQPM}\xspace}
\newcommand{\MCMC}{\ac{MC}\xspace}
\newcommand{\ML}{\ac{ML}\xspace}

%% file: Acknowledgments.tex
%
%

\begin{acknowledgments}
\it 
The dilution refrigerator used for the tests and installed at IJCLab (Orsay, France) was donated by the Dipartimento di Scienza e Alta Tecnologia of the Insubria University (Como, Italy). This work makes use of the \diana data analysis and \apollo data acquisition software which has been developed by the CUORICINO, CUORE, LUCIFER, CUPID-Mo and CUPID-0 collaborations. A.F.L. also acknowledges the support of the California Alliance Fellowship. 
J.K. would like to acknowledge funding from Academy of Finland Grant Nos. 314733 and 345869. This work has also been partially supported by the Academy of Finland under the Academy Project No. 318043. 
F.A.D. and V.I.T. were supported in part by the National Research Foundation of Ukraine Grant No. 2020.02/0011 and would like to acknowledge the heroic efforts of the Armed Forces of Ukraine. 
\end{acknowledgments}